\documentclass[a4,11pt]{reportN}

\usepackage[brazil]{babel}      

\usepackage[latin1]{inputenc}   

\usepackage{graphics}
\usepackage{subfigure}
\usepackage{graphicx}
\usepackage{epsfig}
\usepackage[centertags]{amsmath}
\usepackage{graphicx,indentfirst,amsmath,amsfonts,amssymb,amsthm,newlfont}
\usepackage{longtable}
\usepackage{cite}
\usepackage[usenames,dvipsnames,svgnames,table]{xcolor}
\usepackage{float}
\usepackage[numbers]{natbib}
\usepackage{graphics}
\usepackage{subfigure}
\usepackage{graphicx}
\usepackage{multirow}
\usepackage{epsfig}
\usepackage[centertags]{amsmath}
\usepackage{graphicx,indentfirst,amsmath,amsfonts,amssymb,amsthm,newlfont}

\theoremstyle{remark}

\usepackage{longtable}
\usepackage{cite}
\usepackage{epstopdf}
\usepackage[hidelinks]{hyperref}
\usepackage[usenames,dvipsnames,svgnames,table]{xcolor}
\usepackage{verbatim}

\usepackage[english, ruled, linesnumbered]{algorithm2e}

\usepackage{icomma}

\begin{document}

\title{Note on improvement precision of recursive function simulation in floating point standard
}

\author
    { \Large{Melanie Rodrigues e Silva}\thanks{me\_rodrigues\_silva@hotmail.com} \\
 {\small Group of Control and Modeling, UFSJ, S\~{a}o Jo\~{a}o del-Rei, MG, Brazil}\\
  \Large{Erivelton Geraldo Nepomuceno}\thanks{nepomuceno@ufsj.edu.br} \\
     {\small Department of Electrical Engineering, UFSJ, S\~{a}o Jo\~{a}o del-Rei, MG, Brazil}\\
   \Large{Samir Angelo Milani Martins}\thanks {martins@ufsj.edu.br} \\
 {\small Department of Electrical Engineering, UFSJ, S\~{a}o Jo\~{a}o del-Rei, MG, Brazil}}

\criartitulo


\begin{abstract}
{\bf Abstract}. 
An improvement on precision of recursive function simulation in IEEE floating point standard is presented. It is shown that the average of rounding towards negative infinite and rounding towards positive infinite yields a better result than the usual standard rounding to the nearest in the simulation of recursive functions. In general, the method improves one digit of precision and it has also been useful to avoid divergence from a correct stationary regime in the logistic map. Numerical studies are presented to illustrate the method.


\end{abstract}

\section{Introduction}

The mathematical implementation of functions and systems in computers enables the scientific investigation in many areas. According to \cite{cobem}, interferences and noises of real systems are not incorporated into the simulations, though it is known that it is hardware as a set of real physical systems then the noises are always present. However, limitations of software and hardware are obstacles to the reliability of  results \cite{Gal2013,Loz2013,Nep2014,NM2016,MN2016,NM2017}. Nevertheless, alternatives have been developed to circumvent hardware shortcomings and improve results, such as in \cite{sugihara1989two, wang2014recursive,mcdonough1995new,shi2016algorithms,maharana1997algorithm}. These alternatives that restricts the effects of hardware limitations are usually called as rigorous computing, which is based on the development of refined methods for the implementation of algorithms \cite{Gal2013,Par2012}. 

Chaotic systems such as discrete maps and problems involving recursive functions have a differentiated degree of complexity, since the current iteration depends on the previous ones \cite{Ott2002}. The study of chaotic systems still receive attention and many of scientific conclusions in this area relies upon computer simulation. Chaotic behaviour has been associated to many real applications, from electronic circuits \cite{Mat1984} to ultrasonic cutting system, which dynamic behavior is analyzed using a two-degree-of freedom Duffing oscillator model \cite{yu2016numerical}. Moreover, these iterations are usually evaluated as numerical ill-conditioned \cite{Cor1994}. In general, the simulation process of these types of functions involves errors and its reliability has been questioned in many works \cite{Gal2013,Loz2013}. In this sense, it is well known that in a floating point environment, a computer simulation of a chaotic system may be limited to a short number of iterates \cite{NM2017}. To deal with this problem, many tools have been investigated. We may summarize these tools in two categories. The first is focused on hardware performance using parallel and cluster computation \cite{LW2014}. The second major categories is devoted to improve the algorithm, using interval analysis, significance arithmetic or noisy-mode computation \cite{Par2012}. In this paper, we focus our attention on the noisy-mode computation, but instead of adding pseudo-random digits, as suggested in \cite{Par2012}, we exploit the random nature of rounding error \cite{FDF1992,Gad2015}. 
Considering that the rounding errors are uniformly distributed, we present the background necessary for the understanding of the guidelines and the algorithm of a method that improves the precision of the computational simulation of recursive functions. In order to prove the efficiency of the proposed strategy, we bring examples of the logistic map case.\medskip

\section{Background}
\label{sec:methodology}
This section contains some definitions about recursive functions, orbits, and interval pseudo-orbits. Let $n \in \mathbb{N}$, be a metric space such that $ I \subseteq \mathbb{R} $, and $ f: I \to \mathbb{R} $. Thus, recursive function can be defined as \cite{Nep2014}:
	\begin{equation}
	x_n=f(x_{n-1}),
	\label{eq:recfunction}
	\end{equation}
	or by composite functions:
	\begin{equation}
	x_n=f_1(x_{n-1})=f_2(x_{n-2})=\ldots=f_n(x_{0}),
	\end{equation}	
where $f : I \to  I$ is a recursive function or a map of a state space into itself and $x_n$ denotes the state at the discrete time $n$. As suggested by \citep{gilmore2012topology}, the sequence ${x_n}$ obtained by iterating Eq. \eqref{eq:recfunction} starting from an initial condition $x_0$ is called the orbit of $x_0$. 


\textbf{Definition 1.}\textit{ An orbit is a sequence of values of a map, represented by ${x_n} = [x_0,x_1,...,x_n]$. }

The \textit{Definition 2.} is suggested by \cite{MN2016} and highlights the presence of the difference between real and computational pseudo-orbits.

\textbf{Definition 2.} \textit{Let $i \in \mathbb{N}$ represents a pseudo-orbit, which is defined by an initial condition and a combination of software and hardware. A pseudo-orbit is an approximation of an orbit and can be represented as
 \begin{equation}
 \{\hat{x}_{i,n}\}=[\hat{x}_{i,0},\hat{x}_{i,1},\dots,\hat{x}_{i,n}]   
 \end{equation}
 such that,
 \begin{equation}
 \mid x_n - \hat{x}_{i,n} \mid \le \xi_{i,n}
 \end{equation}
 where $\xi_{i,n} \in \mathbb{R}$ is the limit of error and $\xi_{i,n} \ge 0$.}

There is not an unique pseudo-orbit, as there are different hardware, software, numerical precision standard and discretization schemes, which can produce different results and consequently different errors $\xi_{i,n}$. 

                
%

\section{Method for error reduction}

It is reasonable to assume that round-off errors are uniformly distributed \cite{FDF1992,Gad2015,LPB2007,LN2016}. According to IEEE 754-2008 \cite{IEEE}, the arithmetic operations are also rounded. In this sense, we also assume here that a finite set of arithmetic function presents an error randomly distributed. 

Rounding to the nearest is the most usual for computational arithmetic. However, there are software that enables the user to set the rounding mode towards positive infinite or negative infinite. With these considerations in mind, we establish the main contribution of this letter in the following Lemma.

\textbf{Lemma 1.}\label{lem:lbe} \textit{
        Let $\hat{{x}}_{i,n}^-$ and ${\hat{x}}_{i,n}^+$ be the calculated value by round towards positive infinite and round towards negative infinite, respectively. The arithmetic average given by
            \begin{equation}
                \hat{x}_{j,n}=\frac{\hat{{x}}_{i,n}^+ + {\hat{x}}_{i,n}^-}{2}
                \label{average}
            \end{equation}
      such as $\hat{x}_{j,n} = f(\hat{x}_{j,n-1}) + {\delta}_{j,n}$, presents an round-off error smaller than than the round-off error due the round to nearest as $ n \to \infty $, therefore ${\delta}_{j,n} < {\xi}_{i,n}$.}    

\noindent \textit{Proof.} 
Assuming Eq. \eqref{average}, considering that $\hat{{x}}_{i,n}^- = f(\hat{x}_{j,n-1}) + \delta^-_{i,n}$ and ${\hat{x}}_{i,n}^+ = f(\hat{x}_{j,n-1}) + \delta^+_{i,n}$, then:
\begin{equation}
    \hat{x}_{j,n}=\frac{f(\hat{x}_{j,n-1}) + \delta^-_{i,n} +f(\hat{x}_{j,n-1}) + \delta^+_{i,n}}{2} 
\Rightarrow \hat{x}_{j,n}=\frac{2f(\hat{x}_{j,n-1}) + \delta^-_{i,n} + \delta^+_{i,n}}{2}
\end{equation}
and
\begin{equation}
    \hat{x}_{j,n}=f(\hat{x}_{j,n-1})+\frac{\delta^-_{i,n} + \delta^+_{i,n}}{2}
\end{equation}
But, as we have been considered $\delta^-_{i,n}$ and $\delta^+_{i,n}$ uniformly distributed and it is well known that averaging a random variable leads to a reduction of the noise power in $ n $, which is in this case is $ 2 $. And that completes the proof. \hfill $ \square $

The Algorithm 1 presents a pseudo-code based on the Lemma 1. The round to the nearest is the standard mode. The operators $round^-$ and $round^+$ stands for round towards negative infinite and round towards positive infinite. Where there is no indication of the round mode, one should consider as the round to the nearest. In this paper, we implemented this algorithm in \textit{Matlab R2016a}. The comparison is made with the high precision values provided by the \textit{VPA} toolbox of \textit {Matlab R2016a}. 

	\begin{center}
	\begin{algorithm}	
                    \caption{Simulation of recursive function based on Lemma 1.}
			$\hat{x}_{i,0};$ ~~\textit{\%Initial condition}\\
			$N;$ ~~\textit{\%Number of iterates}\\
			 $\hat{x}^{-}_{i,0} \leftarrow round^-(\hat{x}_{i,0});$\\
			 $\hat{x}^{+}_{i,0} \leftarrow round^+(\hat{x}_{i,0});$\\
            $\hat{x}_{j,0} \leftarrow (\hat{x}^-_{i,0} + \hat{x}^+_{i,0})/2;$ ~~\textit{\%Average of the two round modes}\\
			$\hat{x}^-_{i,0} \leftarrow \hat{x}_{j,0};$\\
			 $\hat{x}^+_{i,0} \leftarrow \hat{x}_{j,0};$\\
			\textbf{For} $n \leftarrow 1$ \textbf{to} $N$ \textit{
			\%Main loop}\\
			$\hat{x}^-_{i,n} \leftarrow round^-(f(\hat{x}^-_{
			j,n-1}));$\\
			$\hat{x}^+_{i,n} \leftarrow round^+(f(\hat{x}^+_{j,n-1}));$\\
			$\hat{x}_{j,n} \leftarrow (\hat{x}^-_{i,n} + \hat{x}^+_{i,n})/2;$\\
			$\hat{x}^-_{i,n} \leftarrow \hat{x}_{j,n};$\\
			 $\hat{x}^+_{i,n} \leftarrow \hat{x}_{j,n};$\\
			 \textbf{EndFor}
	\end{algorithm}

	\end{center}
	
\section{Numerical Examples}

This sections presents three numerical examples using the method proposed in this paper. All the examples are based on the logistic map \cite{May1976}
\begin{equation}
    \label{mp1}
    x_{n+1} = r x_n(1-x_n),
\end{equation}
where the initial condition and the value of bifurcation parameter $r$ are indicated in each example.

\textbf{Example 4.1.}
\textit{Logistic map with $ x_0 = 1 / 3.9 $ and $ r = 3.9 $.}
    
Only using the round to nearest, the pseudo-orbit presents a chaotic behaviour. However, using the Algorithm 1, it is observed that the logistic map function converges to a fixed point as shown in Table \ref{table_conv}. As one can see, the $ \hat{x}_{j,n} $ reaches a fixed point, which is the exact answer for this example, as verified in the following equations: 
\begin{eqnarray}
x_{1} & = & 3.9 (1/3.9)(1- (1/3.9)) = (1 - 10/39) = 29/39 \\
x_{2} & = & 3.9(29/39) (1-29/39) = 29/39
\end{eqnarray} 
and $ 29/39 \approx 0.743589743589744 $ is a fixed point, which is the value that $ \hat{x}_{j,n} $ converges to. On the other hand, the pseudo-orbit $ \hat{x}_{i,n} $ diverges and presents a chaotic behaviour.

    \begin{table}[!ht]
        \doublerulesep 0.1pt
        \tabcolsep 7.8mm
        \centering
        \caption{\rm Iterates of the logistic map with $x_0 = 1/3.9$ and $ r =3.9 $. $ \hat{x}_{i,n} $ is following the rounding to the nearest and $ \hat{x}_{j,n} $ is following the Algorithm 1. The values are shown in decimal and hexadecimal notation. Notice that the third iterate is equal to the second for the Algorithm 1. This can be seen only in the hex format.}	
        \vspace*{1mm}
        \renewcommand{\arraystretch}{1.1}
        \setlength{\tabcolsep}{10pt}
        \footnotesize{\begin{tabular*}{14cm}{ccclcc}
                \hline
        	$ n $ & $ \hat{x}_{i,n} $ (hex) & $ \hat{x}_{i,n} $ (dec) & $ \hat{x}_{j,n} $ (hex) & $ \hat{x}_{j,n} $ (dec) &  \\ \hline
        	  1   &  3fd0690690690691   & 0.256410256410256   & 3fd0690690690691    & 0.256410256410256 \\
        	  2   &  3fe7cb7cb7cb7cb8   & 0.743589743589744   & 3fe7cb7cb7cb7cb8    & 0.743589743589744 \\
        	  3   &  3fe7cb7cb7cb7cb7   & 0.743589743589744   & 3fe7cb7cb7cb7cb8    & 0.743589743589744 \\
        	  4   &  3fe7cb7cb7cb7cb9   & 0.743589743589744   & 3fe7cb7cb7cb7cb8    & 0.743589743589744 \\
        	  5   &  3fe7cb7cb7cb7cb5   & 0.743589743589743   & 3fe7cb7cb7cb7cb8    & 0.743589743589744 \\
        	  6   &  3fe7cb7cb7cb7cbd   & 0.743589743589744   & 3fe7cb7cb7cb7cb8    & 0.743589743589744 \\
        	  7   &  3fe7cb7cb7cb7cae   & 0.743589743589743   & 3fe7cb7cb7cb7cb8    & 0.743589743589744 \\
        	  8   &  3fe7cb7cb7cb7ccb   & 0.743589743589746   & 3fe7cb7cb7cb7cb8    & 0.743589743589744 \\
        	  9   &  3fe7cb7cb7cb7c93   & 0.743589743589740   & 3fe7cb7cb7cb7cb8    & 0.743589743589744 \\
        	 10   &  3fe7cb7cb7cb7cfd   & 0.743589743589751   & 3fe7cb7cb7cb7cb8    & 0.743589743589744 \\ \hline
        \end{tabular*}
        }
        \renewcommand{\arraystretch}{1}
        \label{table_conv}
    \end{table}
 
\textbf{Example 4.2.} \textit{Logistic map with  $ r = 3.9 $ and $ x_0 = 0.01 $.}

The Figure 1 shows the logarithm (base 10) of the error for the first 20 iterates of the logistic map, using the Algorithm 1 and a traditional method based on the round to the nearest. The Algorithm 1 starts with an error greater than the traditional method, but after few iterates the error becomes smaller. The error presented in Figure 1 has been calculated by means of VPA toolbox of Matlab with 1000-digit precision.
    \begin{figure}[!ht]
        \centering
        \label{graf}
        \includegraphics[width=0.7\linewidth]{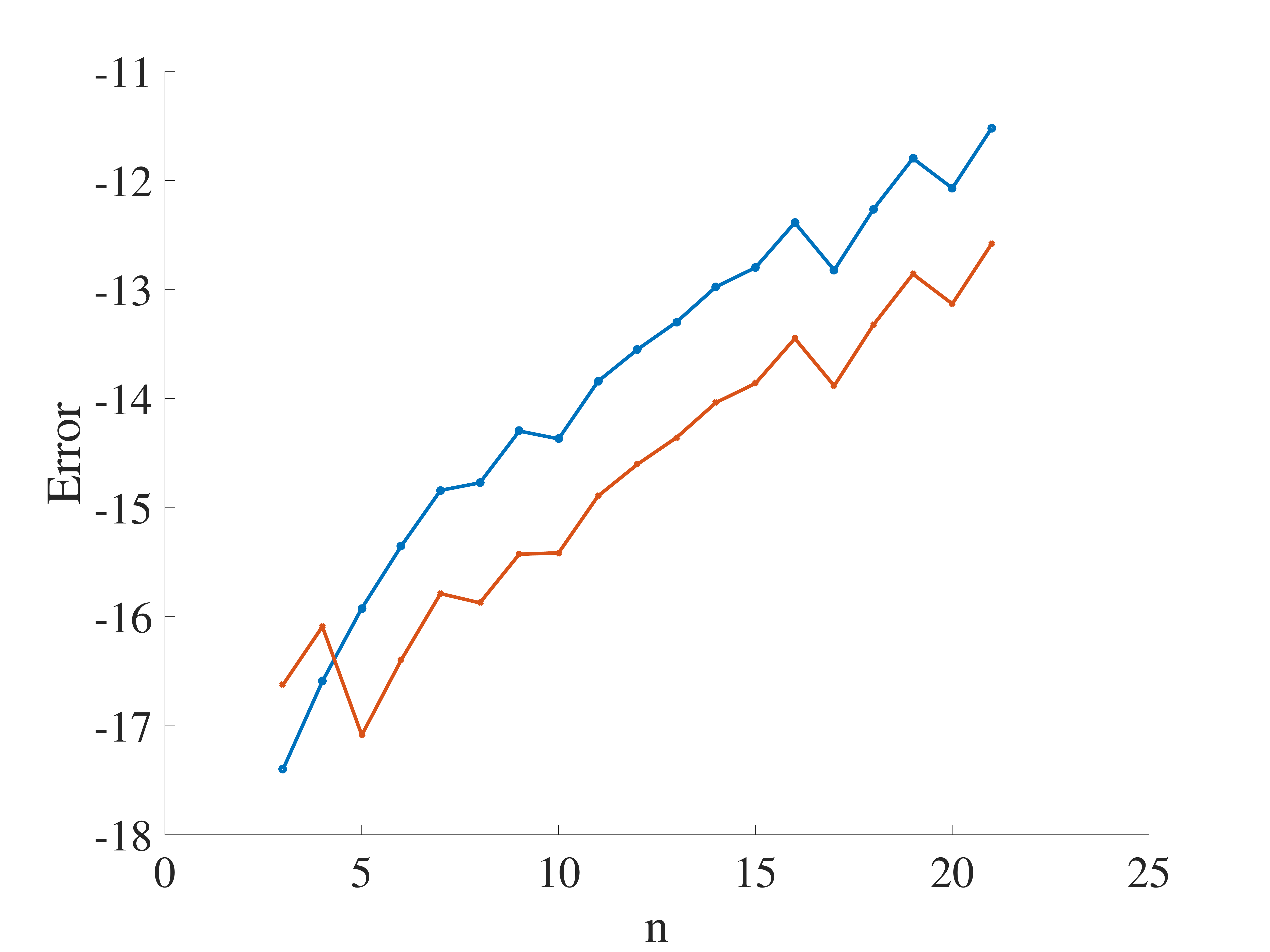}
        \caption{Propagation error for the logistic map with $r=3.9$ and $x_0=0.01$. Traditional method (blue). Algorithm 1 (red)}	
    \end{figure}

\textbf{Example 4.3.} \textit{
Logistic map with  three sets of initial conditions and bifurcation parameter, given by $ (x_0, r) = [(0.1,4.2); (0.2,4);(0.41,3.85)]$.}

Table \ref{table} shows results of $ {\xi}_{i,n} $ and $ \delta_{j,n} $ for the first ten iterations of the logistic map case, considering these three different set of initial conditions and bifurcation parameter. It is possible to note a general decrease in the error that keeps along the iteration process. We arbitrary chosen these parameters, but we have performed our tests with many other sets with similar results.
\label{sec:results}
\begin{table}[!ht]
\centering
\caption{\rm Results of $ \xi_{i,n} $ and $ \delta_{j,n} $ of the logistic map case, considering different initial conditions, $x_0$ and $ r $ values.}
\footnotesize{\begin{tabular*}{14cm}{ccccccc}
\hline
&\multicolumn{2}{c}{${x}_0=0.1$ and $r=4.2$}  &   
 \multicolumn{2}{c}{${x}_0=0.2$ and $r=4$} & 
 \multicolumn{2}{c}{${x}_0=0.41$ and $r=3.85$}\\
\hline
$n$&$\xi_{i,n}$ &$\delta_{j,n}$ &$\xi_{i,n}$ &$\delta_{j,n}$&$\xi_{i,n}$ &$\delta_{j,n}$ \\ \hline
1& 9.33254 e-18&	9.33254 e-18&	4.57088 e-16&	1.20226 e-16& 3.63078 e-16&	2.69153 e-17 \\
			2& 3.98107 e-17&	3.31131 e-17&	7.58577 e-16&	9.12010 e-17& 7.76247 e-16& 6.76082 e-18	\\
			3& 1.38038 e-16&	8.31763 e-17&	1.99526 e-15&	2.08929 e-16& 1.25893 e-15& 4.67735 e-17	\\
			4& 3.38844 e-16&	1.17489 e-17&	1.41253 e-15&	6.76082 e-17& 2.81838 e-15& 1.28824 e-16	\\
			5& 1.86208 e-15&	7.58577 e-16&	5.24807 e-15&	2.69153 e-16& 3.23593 e-15& 2.51188 e-16	\\
			6& 1.90546 e-15&	6.91830 e-16&	1.62181 e-14&	8.51138 e-16& 9.33254 e-15& 7.07945 e-16	\\
			7& 7.76247 e-15&	2.81838 e-15&	1.28824 e-14&	6.76082 e-16& 6.02559 e-15& 4.67735 e-16	\\
			8& 2.88403 e-14&	1.04712 e-14&	4.78630 e-14&	2.51188 e-15& 1.99526 e-14& 1.54881 e-15	\\
			9& 6.30957 e-14&	2.29086 e-14&	1.34896 e-13&	7.07945 e-15& 4.16869 e-14& 3.23593 e-15	\\
			10& 1.41253 e-13&	5.12861 e-14&	4.16869 e-15&	1.94984 e-16& 5.88843 e-14& 4.46683 e-15	\\
				\\\hline\end{tabular*}
		}
		\renewcommand{\arraystretch}{1}
\label{table}
	\end{table}

\section{Conclusions} 
\label{sec:conclusions}
This paper presents a method based on the average of the round modes which promotes an better performance in the simulation of recursive function. The method exploits the stochastic nature of rounding. It consists on the average of two round mode which promotes a reduction of the power noise by a factor of 2. This is a good example of, although a simulation can be seen as a theoretical experiment, in fact, it presents some real world dimension, as stochasticity, and therefore, usual tools presented in Engineering can be applied. 

We applied the method in three numerical examples. In the first case, we show that the method is able to keep the correct fixed point, while the rounding to the nearest diverge to chaotic behaviour. In the second and third we show the general behaviour of the method in reducing the error of the simulation. 

It is important to say that there is no significant increase of computational time and this method can be easily extended to many other applications, such as numerical solution of differential equations, as seen applied in \cite{yu2016numerical}. In the future, we intend to investigate the use of interval extensions and other rounding modes as further discussed in \cite{NM2017,MN2016}.

\subsection*{Acknowledgements}

This work has been supported by the CNPq and CAPES. A special thanks to Control and Modelling Group (GCOM).

\end{document}